\documentclass[runningheads]{llncs}
\usepackage{graphicx}
\usepackage{multirow}
\usepackage{tabularx}
\usepackage[normalem]{ulem}
\useunder{\uline}{\ul}{}
\usepackage{xcolor}
\usepackage{hyperref}
\usepackage{bm}
\usepackage{marvosym}

\begin{document}
\title{Groupwise Query Performance Prediction with BERT}
\titlerunning{Groupwise Query Performance Prediction with BERT}
\author{Xiaoyang Chen\inst{1,2} \and
Ben He\inst{1,2}\textsuperscript{\Letter} \and
Le Sun\inst{2}}
% %
\authorrunning{X. Chen et al.}
\institute{University of Chinese Academy of Sciences, Beijing, China \and
Institute of Software, Chinese Academy of Sciences, Beijing, China \\
\email{chenxiaoyang19@mails.ucas.ac.cn}\\
\email{benhe@ucas.ac.cn}, \email{sunle@iscas.ac.cn}
}
\maketitle              % typeset the header of the contribution
\begin{abstract}
%BERT在多种IR任务中取得sota效果，但在QPP任务中的应用还几乎是空白，尤其是在传统TREC数据集上。
%在本文中，我们将BERT应用于QPP任务，并在此基础上尝试进一步结合在检索任务中被证明有效的groupwise方法。
%基于检索模型和初始QPP模型的结果，文档被分为不同的group，这些group包含inter-query级别或intra-query级别上下文。
%在三个TREC数据集上的结果表明，通过使用恰当的监督信号，在少量数据上训练的BERT就能够在QPP任务上取得与目前已有方法相近的效果，而groupwise方法能够进一步提升BERT效果
While large-scale pre-trained language models like BERT have advanced the state-of-the-art in IR, its application in query performance prediction (QPP) is so far based on pointwise modeling of individual queries. %has not been fully explored.
% a largely underexplored domain.
Meanwhile, recent studies suggest that the cross-attention modeling of a group of documents can effectively boost performances for both learning-to-rank algorithms and BERT-based re-ranking. To this end, a BERT-based groupwise QPP model is proposed, in which the ranking contexts of a list of queries are jointly modeled to predict the relative performance of individual queries.
%According to the results of the retrieval model and an existing QPP method, the documents are divided into different groups that contain inter-query or intra-query ranking contexts.
Extensive experiments on three standard TREC collections showcase effectiveness of our approach. Our code is available at \url{https://github.com/VerdureChen/Group-QPP}.
%that by using proper supervision signals, the proposed model achieves promising results with low training and test computation cost.
%Impact of different factors, including the batch size of training and test, model size, and ranking context of queries, on the model prediction accuracy are also investigated.

% the proposed model outperforms state-of-the-art baselines with low training and test computation cost.
% BERT achieves comparable results to existing metrics in the literature, and its effectiveness can be further improved by combining the groupwise method.

% \keywords{First keyword  \and Second keyword \and Another keyword.}
\end{abstract}
\section{Introduction}\label{sec.introduction}
%first paragraph, introduce the task and the background.
%second paragraph, what is needed for the task.
%We propose a method by this thought, give an introduction of our method.
%The novelty and contributions.
%What to explore in this paper.

Query performance prediction (QPP) aims to automatically estimate the search results quality of a given query. While the pre-retrieval predictors enjoy the low computational overhead \cite{DBLP:conf/sigir/Cronen-TownsendZC02,DBLP:journals/is/HeO06,DBLP:conf/ecir/HeLR08,Mothe2005LinguisticFT}, the post-retrieval methods are in general more effective by considering sophisticated query and document features \cite{DBLP:conf/ecir/ArabzadehBZB21,DBLP:conf/ecir/AslamP07,DBLP:conf/sigir/Cronen-TownsendZC02,DBLP:conf/sigir/CumminsJO11,DBLP:conf/sigir/Diaz07,DBLP:conf/spire/Perez-IglesiasA10,DBLP:conf/ictir/RoitmanEW17,DBLP:conf/ictir/ShtokKC09,DBLP:conf/cikm/TaoW14,DBLP:conf/sigir/VinayCMW06,DBLP:conf/sigir/ZamaniCC18,DBLP:conf/cikm/ZhouC06,DBLP:conf/sigir/ZhouC07}. Recently, the large-scale pre-trained transformer based language models, e.g. BERT~\cite{DBLP:conf/naacl/DevlinCLT19}, has shown to advance the ranking performance, which provides a new direction for task of QPP. 

Indeed, recent results demonstrate that BERT effectively improves the performance of post-retrieval QPP~\cite{web/bertqpp,DBLP:conf/ictir/HashemiZC19}. For instance, training with a large number of sparse-labeled queries and their highest-ranked documents, BERT-QPP~\cite{web/bertqpp} examines the effectiveness of BERT on the MS MARCO~\cite{DBLP:journals/corr/NguyenRSGTMD16} and TREC DL~\cite{DBLP:journals/corr/abs-2102-07662,DBLP:journals/corr/abs-2003-07820} datasets, by pointwise modeling of query-document pairs. 
Beyond learning from single query-document pairs, the groupwise methods have achieved superior performance on both learning-to-rank \cite{DBLP:conf/sigir/AiBGC18,DBLP:conf/ictir/AiWBGBN19,DBLP:journals/corr/abs-1910-09676,DBLP:conf/sigir/PangXALCW20} and BERT re-ranking~\cite{DBLP:journals/corr/abs-2104-08523} benchmarks.
To this end, we propose an end-to-end BERT-based QPP model, which employs a groupwise predictor to jointly learn from multiple queries and documents, by incorporating both cross-query and cross-document information.
Experiments conducted on three standard TREC collections show that our model improves significantly over state-of-the-art baselines.

\section{Related Work}\label{sec.related_work}

\noindent{\bf Query performance prediction (QPP).} 
Early research in QPP utilizes linguistic information~\cite{Mothe2005LinguisticFT}, statistical features~\cite{DBLP:conf/sigir/Cronen-TownsendZC02,DBLP:journals/is/HeO06,DBLP:conf/ecir/HeLR08} in pre-retrieval methods, or analyses clarity~\cite{DBLP:conf/sigir/Cronen-TownsendZC02,DBLP:journals/ir/Cronen-TownsendZC06}, robustness~\cite{DBLP:conf/ecir/AslamP07,DBLP:conf/sigir/Diaz07,DBLP:conf/sigir/VinayCMW06,DBLP:conf/cikm/ZhouC06,DBLP:conf/sigir/ZhouC07}, retrieval scores~\cite{DBLP:conf/spire/Perez-IglesiasA10,DBLP:conf/ictir/RoitmanEW17,DBLP:conf/ictir/ShtokKC09,DBLP:conf/cikm/TaoW14,DBLP:conf/sigir/ZhouC07} for post-retrieval 
prediction, which further evolves into several effective frameworks~\cite{DBLP:conf/sigir/CumminsJO11,DBLP:conf/sigir/Diaz07,DBLP:conf/ictir/KurlandSCH11,DBLP:conf/sigir/Roitman17,DBLP:conf/ictir/RoitmanESW17,DBLP:conf/sigir/ShtokKC10,DBLP:journals/tois/ShtokKC16}. The QPP techniques have also been explored and analyzed in~\cite{DBLP:conf/ecir/ArabzadehBZB21,DBLP:journals/ipm/ArabzadehZJAB20,DBLP:conf/ecir/ArabzadehZJB20,DBLP:conf/sigir/ChifuLMU18,DBLP:conf/sac/DejeanIMU20,DBLP:conf/ecir/FaggioliZCFS21,DBLP:conf/ictir/HashemiZC19,DBLP:journals/ipm/KhodabakhshB21,DBLP:conf/sigir/RaiberK14,DBLP:conf/sigir/RavivKC14,DBLP:conf/ictir/Roitman20,DBLP:conf/sigir/RoitmanK19,DBLP:conf/ictir/RoitmanMFS20,DBLP:conf/sigir/ZendelCS21,DBLP:conf/sigir/ZendelSRKC19,DBLP:conf/cikm/KrikonCK12}. 
% Arabzadeh et al. propose to measure the coherence over the document association network of individual queries~\cite{DBLP:conf/ecir/ArabzadehBZB21}.
With the recent development deep learning techniques, NeuralQPP~\cite{DBLP:conf/sigir/ZamaniCC18} achieves promising results by training a three-components deep network under weak supervision of existing methods. 
Recently, %large-scale pre-trained transformer-based language models like BERT~\cite{DBLP:conf/naacl/DevlinCLT19} has advanced the state-of-the-art of information retrieval. 
while NQA-QPP~\cite{DBLP:conf/ictir/HashemiZC19} uses BERT to generate contextualized embedding for QPP in non-factoid question answering, BERT-QPP~\cite{web/bertqpp} directly applies BERT with pointwise learning in the prediction task, outperforming previous methods on the MS MARCO dev set~\cite{DBLP:journals/corr/NguyenRSGTMD16} and TREC Deep Learning track query sets~\cite{DBLP:journals/corr/abs-2102-07662,DBLP:journals/corr/abs-2003-07820}.% Despite the success of BERT-QPP, it is trained in a pointwise manner, ignoring the relative relations among queries and documents. 
% taking only one document for the judgment of each query makes how to adapt it to the standard TREC collections with a small number of queries (such as robust04) a problem to be explored.

\noindent{\bf Groupwise Ranking.} %A considerable amount of work has attempted to incorporate the idea of groupwise into ranking tasks. 
Beyond pointwise loss, pairwise and listwise losses are proposed to learn the relative relationships among  documents~\cite{DBLP:journals/corr/abs-2108-03586}. Recently, Ai et al. ~\cite{DBLP:conf/sigir/AiBGC18} propose to represent documents into embedding with an RNN and refine the rank lists with local ranking context. 
Thereafter, a groupwise scoring function is proposed by Ai et al. ~\cite{DBLP:conf/ictir/AiWBGBN19} to model documents jointly. 
In the learning-to-rank context, both Pasumarthi et al.~\cite{DBLP:journals/corr/abs-1910-09676} and Pang et al.~\cite{DBLP:conf/sigir/PangXALCW20} use self-attention mechanism with groupwise design to improve retrieval effectiveness.
% Finding out current BERT-based re-ranking models score each query-document pair independently without consider other documents in the same list,  
Furthermore, Co-BERT~\cite{DBLP:journals/corr/abs-2104-08523} incorporates cross-document ranking context into BERT-based re-ranking models, demonstrating the effectiveness of using groupwise methods in boosting the ranking performance of BERT. 
In brief, while previous works are carried out on single query-document pairs with BERT, the groupwise methods have shown useful in multiple studies. To this end, this work proposes a groupwise post-retrieval QPP model based on pre-trained language models which simultaneously takes multiple queries and documents into account.

\section{Method}\label{sec.method}

\begin{figure}[!t]
\centering
\includegraphics[width=0.9\textwidth]{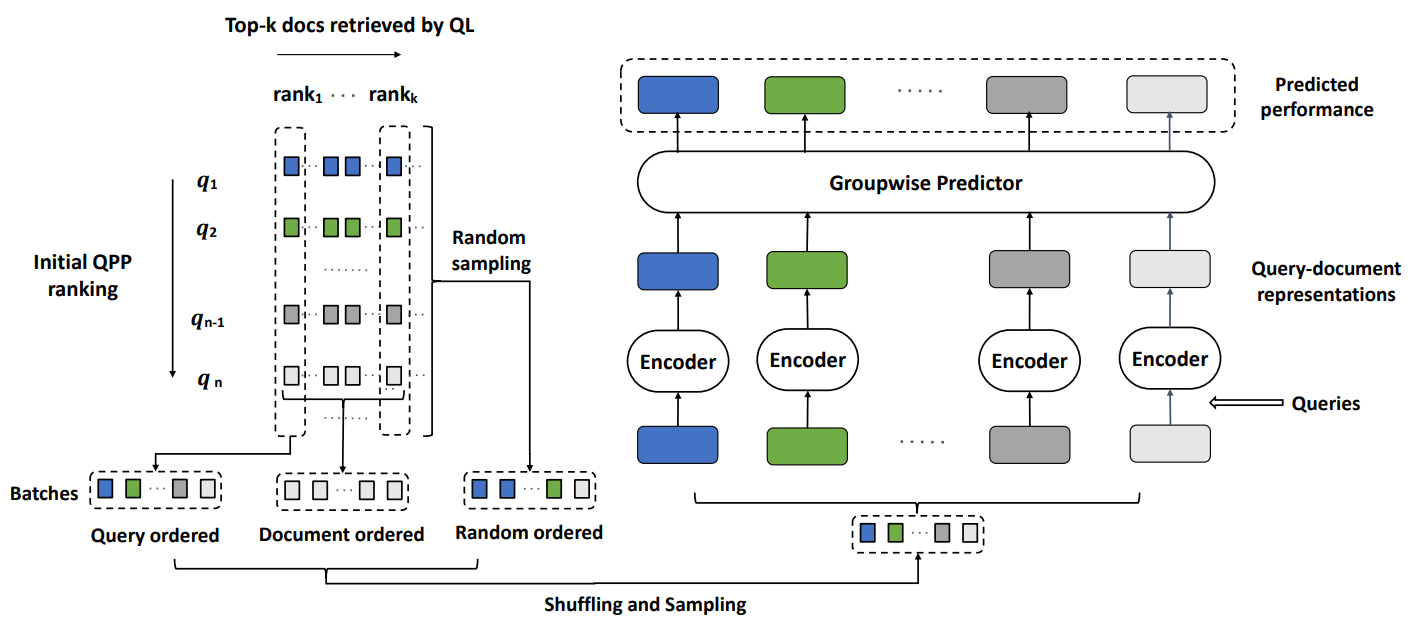}
\caption{Model architecture of the proposed groupwise framework. }
\label{fig.model_architecture}
\end{figure}

Figure~\ref{fig.model_architecture} shows our model architecture. Give an underlying retrieval method $M$ and a corpus $C$, in response to a query $q$, a document set $D$ is composed by the top$_k$ documents retrieved from $C$ with $M$.
As aforementioned, existing BERT-based QPP methods only use the text from individual query-document pairs; 
however, considering information from different queries and documents is necessary for QPP tasks,
which aim to obtain relative performance among queries.
Inspired by Co-BERT~\cite{DBLP:journals/corr/abs-2104-08523}, to boost the performance of BERT-based QPP methods, a groupwise predictor is integrated to learn from multiple queries and documents simultaneously on the basis of a BERT encoder.

\noindent{\bf Encoding query-document pairs.} Following Arabzadeh et al.~\cite{web/bertqpp}, we first encode each query-document pair with BERT. As documents are frequently long enough to exceed BERT's 512 token limit, similar to Co-BERT~\cite{DBLP:journals/corr/abs-2104-08523}, we split long texts into equal-sized passages. We use a BERT checkpoint fine-tuned on MS MARCO~\cite{DBLP:journals/corr/NguyenRSGTMD16} to predict the relevance between each query and its corresponding passages. Each document used in the next steps is represented by its top-1 ranked passage.
% For brevity, we use the term document, represented by the selected passage, in the following.
Consistent with common practices for text categorization using BERT, the token sequences $[CLS]Query[SEP]Document[SEP]$ are put into BERT to get encoded. %The representation for $[CLS]$ which encodes the interaction information of the query and the document is subsequently passed to the full connection layer to get the predicted score.   
We use the $[CLS]$ representation in the following groupwise step to further integrate the cross-query as well as cross-document information.

\noindent\textbf{Groupwise predictor.} To incorporate cross-document and cross-query context, we regard each batch as a single group of query-document pairs. Suppose the batch size is $n$ ($n\leq k$), the $[CLS]$ vectors in a batch are reshaped into a sequence of length $n$, and we denote the sequence as $z_{1},z_{2},z_{3},\cdots, z_{n}$. For $i\in [1,\cdots ,n]$, each $z_i$ is a $d$-dimensional vector, for example, $d=768$ for BERT-Base. 
Similar to Chen et al.~\cite{DBLP:journals/corr/abs-2104-08523}, we use a four-layers transformer as the groupwise predictor, which enables the cross attention among the $[CLS]$ vectors in each batch, and then produces $n$ predicted performances of each query-document pair.
During inference, suppose top$_t$ documents of $q$ are used, we will get $t$ predicted scores for $q$. We use three aggregation methods to get the final QPP score of $q$: max-pooling, average-pooling, and the direct use of the predicted performation of the first-ranked retrieved document for query $q$. In our experiments, the aggregation method with the best performance on the training set is chosen.

By assigning different positional ids to $z_i$, our model can be designed to incorporate with different types of ranking context.
Thus, several \textbf{variants of our models} are investigated. \textbf{(Random order)} denotes that all query-document pairs are shuffled before being fed into the model in both training and inference.
\textbf{(Query order)} denotes for BERT groupwise model considering only the \textit{cross-query} context. For a batch of $n$ samples, the $i$th ranked documents from $n$ queries are grouped together in the batch, and position ids are assigned by the initial query order derived by $n(\sigma_{X\%})$. We leave other choices of the initial QPP for future study. 
\textbf{(Doc order)}
denotes for BERT groupwise model considering only the \textit{cross-document} context. A batch consists of $n$ documents returned for a query, and the position ids are assigned by the initial document ranking.
\textbf{(Query+Doc)}
denotes for BERT groupwise model considering both cross-document and cross-query context. 
Batches containing one of the above two contexts appear randomly during training.
% Top documents from a list of ordered queries are grouped in a batch.
\noindent\textbf{(R+Q+D)}
denotes for BERT groupwise model with all three types of orders mentioned above.
According to the maximum batch size allowed by the hardware, we use the batch size of 128/64/16 for Small, Base and Large BERT models, respectively. Note that the training data is still shuffled among batches to avoid overfitting.

\section{Experiment Setup}\label{sec.exp_setup}

% \subsection{Dataset and Metrics}%\label{sec.dataset_exp}
\textbf{Dataset and Metrics.} We use three popular datasets, namely, Robust04~\cite{DBLP:conf/trec/Voorhees04b}, GOV2~\cite{DBLP:conf/trec/ClarkeCS04}, and ClueWeb09-B~\cite{DBLP:conf/trec/ClarkeCS09}, with 249, 150 and 200 keyword queries, respectively. Following~\cite{DBLP:conf/ecir/ArabzadehBZB21}, we use the Pearson's $\rho$ and Kendall's $\tau$ correlations to measure the QPP performance, which is computed using the predicted ordering of the queries with the actual ordering of average precision for the top 1000 documents (AP@1000) per query retrieved by the Query Likelihood (QL) model implemented in Anserini~\cite{DBLP:conf/sigir/Yang0L17}. Following~\cite{DBLP:conf/sigir/ZamaniCC18}, we use 2-fold cross-validation and randomly generate 30 splits for each dataset. Each split has two folds, the first fold is used for model training and hyper-parameter tuning% to choose aggregation methods and interpolation weights with the highest Pearson's $\rho$
. The ultimate performance is the average prediction quality on the second test folds over the 30 splits.
Statistical significance for paired two-tailed t-test is reported.

% \subsection{Baselines and Variants}%\label{sec.baselines}
\noindent\textbf{Baselines.} Akin to~\cite{DBLP:conf/ecir/ArabzadehBZB21}, we compare our model with several popular baselines including 
\textbf{Clarity~\cite{DBLP:conf/sigir/Cronen-TownsendZC02}},
\textbf{ Query Feedback (QF)~\cite{DBLP:conf/sigir/ZhouC07}}, 
\textbf{Weighted Information Gain (WIG)~\cite{DBLP:conf/sigir/ZhouC07}}, 
\textbf{Normalized Query Commitment (NQC)~\cite{DBLP:conf/ictir/ShtokKC09}}, 
\textbf{Score Magnitude and Variance (SMV)~\cite{DBLP:conf/cikm/TaoW14}}, 
\textbf{Utility Estimation Framework (UEF)~\cite{DBLP:conf/sigir/ShtokKC10}},
\bm{$\sigma_{k}$}~\cite{DBLP:conf/spire/Perez-IglesiasA10},
\bm{$n(\sigma_{X\%})$}~\cite{DBLP:conf/sigir/CumminsJO11},
\textbf{Robust Standard Deviation (RSD)~\cite{DBLP:conf/ictir/RoitmanEW17}},
\textbf{WAND[\bm{$n(\sigma_{X\%})$}]~\cite{DBLP:conf/ecir/ArabzadehBZB21}},
and \textbf{NeuralQPP~\cite{DBLP:conf/sigir/ZamaniCC18}}. 
\noindent We also compare to \textbf{BERT-Small/Base/Large~\cite{google_url}} baselines, which are configured the same as our model except that they do not have a groupwise predictor. Note that the BERT baselines share the same structures with BERT-QPP except we use more documents for each query in training due to the small number of queries. Following~\cite{DBLP:conf/ecir/ArabzadehBZB21}, our proposed predictor is linearly combined with $n( \sigma_{X\%})$. The BERT baselines perform the same linear interpolation. %All the results are derived from the same cross-validation method as our models.

\noindent\textbf{Data preparation and Model training.} 
% The Prediction quality is measured using coefficients between the ground truth AP@1000 and the values predicted by the proposed model. 
Akin to~\cite{DBLP:journals/corr/abs-2104-08523}, for the BERT-based models, documents are sliced using sliding windows of 150 words with an overlap of 75 words.
The max sequence length of the concatenated query-document pair is 256.
We use MSE loss for individual documents and explore two kinds of training labels: P@k and AP@1000.
\textcolor{black}{According to our pilot study on the BERT-Base baseline, we use P@k as the supervision signals on Robust04 and GOV2, and use AP@1000 on ClueWeb09-B.}
%The BERT-based models are trained with a batch size $n$ of 64 on one GeForce RTX 3090 24G GPU.
%We train the BERT baselines and groupwise models with single order for five epochs to avoid overfitting, and groupwise models with multiple orders are trained for six epochs. 
All BERT models are trained for 5 epochs. Due to the memory limit, BERT-based models are trained with top-100 documents and tested on the last checkpoint with the top-25 documents for each query retrieved by QL.
We use Adam optimizer~\cite{DBLP:journals/corr/KingmaB14} with the learning rate schedule from~\cite{DBLP:journals/corr/abs-1901-04085}. We select the initial learning rate from \{1e-4, 1e-5, 1e-6\}, and set the warming up steps to 10\% of the total steps.

\section{Results}\label{sec.results}

\begin{table*}[!ht]
\centering
\caption{ Evaluation results. Statistical significance at 0.05 relative to BERT baselines
of the same model size (e.g. (R+Q+D)-Large vs. BERT-Large) is marked with *.}
\label{tab:effectiveness}
\resizebox{0.8\textwidth}{!}{%
\begin{tabular}{|l|llllll|}
\hline
\multicolumn{1}{|c|}{\multirow{2}{*}{Method}}  & \multicolumn{2}{l}{Robust04} & \multicolumn{2}{l}{GOV2} & \multicolumn{2}{l|}{ClueWeb09-B} \\ \cline{2-7} 
\multicolumn{1}{|c|}{}                        & P-$\rho$       & K-$\tau$       & P-$\rho$       & K-$\tau$     & P-$\rho$       & K-$\tau$      \\ \hline
Clarity                                       & 0.528          & 0.385          & 0.428          & 0.291          & 0.300          & 0.213        \\
QF                                            & 0.390          & 0.324          & 0.447          & 0.314          & 0.163          & 0.072          \\
WIG                                           & 0.546          & 0.379          & 0.502          & 0.346          & 0.316          & 0.210          \\
NQC                                           & 0.516          & 0.388          & 0.381          & 0.323          & 0.127          & 0.138          \\
SMV                                           & 0.534          & 0.378          & 0.352          & 0.303          & 0.236          & 0.183          \\
UEF                                           & 0.502          & 0.402          & 0.470          & 0.329          & 0.301          & 0.211          \\
$\sigma_{k}$                                  & 0.522          & 0.389          & 0.381          & 0.323          & 0.234          & 0.177          \\
$n(\sigma_{X\%})$                            & 0.589          & 0.386          & 0.556          & 0.386          & 0.334          & 0.247   \\
RSD                                           & 0.455          & 0.352          & 0.444          & 0.276          & 0.193          & 0.096          \\
WAND[$n(\sigma_{X\%})$]                      & 0.566          & 0.386          & 0.580  & 0.411  & 0.236         & 0.142          \\
NeuralQPP                                    & 0.611 &  0.408  & 0.540          & 0.357          &  0.367  & 0.229          \\\hline
BERT-Small                            & 0.591          & 0.391          & 0.615          & 0.436          & 0.394          & 0.278          \\
BERT-Base                             & 0.585          & 0.423          & 0.637          & 0.454          & 0.447          & 0.321 \\
BERT-Large                            & 0.579         &  0.422          & 0.645          &  0.461          & 0.342           &  0.251       \\ \hline
% \multicolumn{7}{|l|}{BERT-Base }                                                                                                                    \\ \hline
% (P@k\_1000)                                & 0.559          & 0.401          & \textbf{0.645}    & \textbf{0.475}  & 0.350          & 0.233          \\
% (AP@1k\_1000)                              & 0.346          & 0.231          & 0.151          & 0.081        & \textbf{0.476}    & \textbf{0.345}          \\
% (P@k\_100)                                 & \textbf{0.585}   & \textbf{0.423}     & 0.637    & 0.454          & 0.393          & 0.271          \\
% (AP@1k\_100)                               & 0.520          & 0.360          & 0.505          & 0.345          & 0.447          & 0.321          \\ \hline

% (Random\_100\_Small)                            &                &                &                &                & 0.443          & 0.306          \\ 

(Random order)-base                            & 0.608*          & 0.449*         & 0.665*          & 0.479*          &  0.481* &  0.353* \\
(Query order)-base                             & \textbf{0.615*} & 0.456*          & 0.676*          & 0.486*          & 0.455          &  0.327          \\
(Doc order)-base                               & 0.563          & 0.383          & 0.660*          & 0.476*          & 0.365          &  0.262         \\
(Query+Doc)-base                               & 0.598          & 0.452*          &  0.682* &  0.496* & 0.438          &  0.317         \\
(R+Q+D)-small                      & 0.590          & 0.419*          & 0.680*          & 0.500*          & 0.437*          & 0.305*         \\
(R+Q+D)-base                                   & 0.608*          &  0.460* & 0.676*          & 0.489*          & 0.449          & 0.324          \\ 
(R+Q+D)-large                             & 0.612*          & \textbf{0.470*} & \textbf{0.688*} & \textbf{0.508*} & \textbf{0.545*} &\textbf{0.399*}    \\
% (Random\_100\_Large)                   &                &                &                &                & 0.531        &  0.396        \\ 
\hline

\end{tabular} 
}
\end{table*}

\noindent\textbf{Overall effectiveness.} According to Table~\ref{tab:effectiveness}, the proposed model outperforms all the baselines on all three collections.
Compared with the previous state-of-the-art results without using BERT, except for the $\rho$ on Robust04, our groupwise model trained on BERT-Base with the random input order has an improvement on all metrics by at least 10\%.
In general, our (R+Q+D) outperforms the BERT baselines with all three different model sizes.
Additionally, varying the type of ranking contexts incorporated with the groupwise models leads to different observations on the three datasets.
The query-level ranking context marginally improves the effectiveness on Robust04 and GOV2, while it decreases the result on CluWeb09-B.
Using document-level ranking context alone greatly harms the model performance on Robust04 and ClueWeb09-B.
This may be due to the fact that the model has only learned the sequence information inside each query, but not the relative relations between the queries.
Relative to the random case, using both contexts slightly elevates the performance on GOV2, while it has little effect on Robust04 and decreases the results on ClueWeb09-B.
As the simultaneous use of all three types of context appears to be the best variant, we only report results of (R+Q+D)-Base in the following analysis.
%For the best of both worlds, we will use three kinds of context information simultaneously in the next experiments and analysis.

\begin{figure}
\centering
\includegraphics[width=0.8\textwidth]{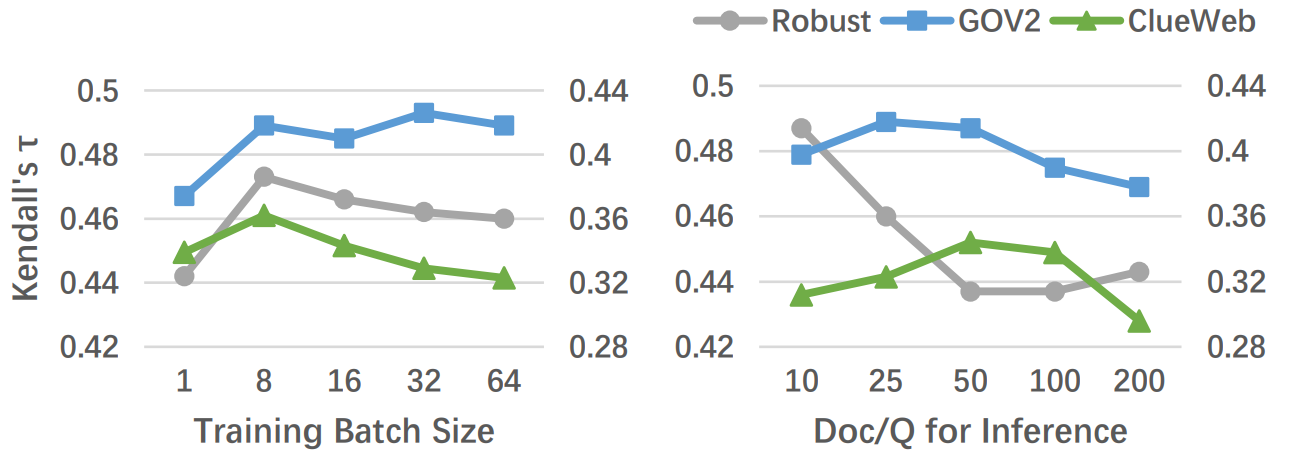}
\caption{Performance of (R+Q+D) with different training batch sizes and numbers of documents per query for inference. In each figure, the left axis represents the Kendall's ${\tau}$ of Robust04 and GOV2, and the right axis represents ClueWeb09-B.  
}
\label{fig.ablation}
\end{figure}

% We then examine the effect of model size and \textcolor{red}{batch size in training and testing} on groupwise.
\noindent\textbf{Impact of factors.} We examine the impact of training batch size and the number of top-$k$ documents per query for inference on the model performance. 
We first evaluate with different training batch sizes in $\{1, 8, 16, 32, 64\}$. The greater the batch size is, the more query-document pairs are jointly modeled. A special case is to set batch size to 1, which is equivalent to the pointwise learning without any context from other queries or documents. The results in Figure~\ref{fig.ablation} show that the cross-attention among queries is effective and improves upon the pointwise method by a large margin. The groupwise method works best with a group size of 8, which means the model may learn better with a relatively smaller group of queries.
We also explore the impact of different numbers of documents per query used during inference, namely $\{10, 25, 50, 100, 200\}$. Results in Figure~\ref{fig.ablation} indicate that inference with less than 100 documents per query on all three collections yields the best results. 
The reason might be that there are more positive samples in the top-ranked documents which contribute more to the target metric, i.e. AP@1000, while considering more negative examples not only have little impact on the target metric, but also are more likely to introduce noise.

\noindent\textbf{Limitations.} 
We count the number of floating-point operations for all BERT-based models. It turns out our model can predict the retrieval performance with less than 1\% additional computational cost compared to its BERT counterpart. However, for document retrieval with BERT MaxP, the passage selection brings an approx. 1 min extra computational overhead, which is more expensive than the non-BERT baselines.

\section{Conclusion}\label{sec.conclusion}
%Although it is well known that referring to other queries to determine the performance of a particular query can lead to more accurate results, this information has not been explicitly modeled in the previous BERT-based QPP models.
In this paper, we propose a BERT-based groupwise query performance prediction method, which simultaneously incorporates the cross-query and cross-document information within an end-to-end learning framework.
Evaluation on three standard TREC test collections indicates 
the groupwise model significantly outperforms the BERT baselines nearly in all cases.
In further research, we plan to work on the efficiency, as well as adoption of our approach to more advanced experimentation framework~\cite{DBLP:conf/ecir/FaggioliZCFS21}.
\newpage
\bibliographystyle{splncs04}
\bibliography{ref}
\end{document}